\documentclass[a4paper,aps,preprintnumbers,twocolumn,amsmath,amssymb,prl,showpacs]{revtex4}
\usepackage{bbm}
\usepackage{mathrsfs}
\usepackage{amsmath}
\usepackage[dvips]{graphicx}
\usepackage{epsfig}
\usepackage[dvips]{graphics,graphicx}

\newcommand{\be}{\begin{equation}}
\newcommand{\ee}{\end{equation}}
\newcommand{\bea}{\begin{eqnarray}}
\newcommand{\eea}{\end{eqnarray}}
\newcommand{\bes}{\begin{split}}
\newcommand{\ees}{\end{split}}

\renewcommand{\vec}[1]{\mathbf{#1}}

\newcommand{\tr}{\operatorname{Tr}}

\usepackage[dvips]{graphics,graphicx}

\begin{document}
\title{Edge State, Entanglement Entropy Spectra and Critical Hopping Coupling of Anisotropic Honeycomb Lattice}
\author{Ming-Chiang Chung$^{1,2}$, Yi-Hao Jhu$^3$, Pochung Chen$^3$ and Sungkit Yip$^2$}
\affiliation{$^1$Physics Division, National Center for Theoretical Science, Hsinchu, 30013, Taiwan}
\affiliation{$^2$Institute of Physics, Academia Sinica, Taipei 11529, Taiwan}
\affiliation{$^3$Physics Department, National Tsing Hua University, Hsinchu, 30013, Taiwan}

\begin{abstract}
  For a bipartite honeycomb lattice, we show that
the Berry phase depends not only on the shape of the system but also on the hopping couplings. Using the entanglement entropy spectra obtained by diagonalizing the block Green's function matrices, 
the maximal entangled state with the eigenvalue $\lambda_m=1/2$ of the reduced density matrix is shown to have one-to-one correspondence to the zero energy states of the lattice with open boundaries, which depends on the Berry phase. For the  systems with finite bearded edges along $x$-direction we find critical hopping couplings: the maximal entangled states (zero-energy states) appear pair by pair if one increases the hopping coupling $h$ over the critical couplings $h_c$s.   
\end{abstract}


\date{\today}
\maketitle

%%%%%%%%%%%%%%%%%%%%%%%%%%%%%%%%%%%%%%%%%%%%%%%%%%%%%%%%%%%%%%%%

%\begin{figure}
%\center
%\includegraphics[width=7.0cm]{fig1.eps}
%\caption{(Color online) Order parameter of the nematic state vs. temperature in unit of $k_B T_N=6|K|$.  Solid line is the mean-field result, while the red dashed line represents the spin-wave approach. The two approximations using linear and quadratic dispersion are shown in  blue dotted line and  green dotted-dashed line, respectively.} \label{fig1}
%\end{figure} 

Experimental studies on graphene\cite{Graphene} and cold atoms\cite{Coldatoms} have inspired new interests in quantum properties of electrons and atoms on the honeycomb lattice. Due to its peculiar energy dispersion, which contains two Dirac points as Fermi surfaces, the honeycomb lattice offers particular physical phenomena which can not be observed on the square lattice. In particular  the edge states with zero energy may appear when  special boundaries are  present. The existence of the edges state relies on the shape of boundaries, that is, for zigzag and bearded edges there exist zero-energy  edge states, on the contrary the armchair edges offer no zero-energy edge modes according to their Berry phase of the occupied band \cite{RyuHatsugai02}. Therefore the topological structure of the system strongly influences the existence of zero-energy edge modes.

On the other hand, recent  development on quantum information applied in condensed matter theory provides a new tool to investigate quantum phenomena \cite{Review}. Especially the entanglement entropy:
$S_A = -\tr \rho_A \log_{2} \rho_A,$ where $ \rho_A=\tr_B |\Psi_{AB}\rangle\langle\Psi_{AB}|$,   
has been widely used to measure the bipartite  entanglement for a given pure state (ground state) $|\Psi_{AB}\rangle$ of a bipartite $AB$ consisting of system $A$ and environment $B$.  In the last several years this quantity  has been successfully used to interpret the nature of the quantum criticality \cite{QCP} due to the fact that the entanglement entropy diverges at the quantum critical points of the second order phase transition. The scaling law of the entanglement entropy even provides a rule to tell different quantum phases \cite{AreaLaw}.  Furthermore, the entanglement entropy spectrum for free fermions, defined as the set of $\lambda_m$, where
 $\rho_A = \bigotimes_{m} \left[\begin{matrix}  \lambda_m & 0\\ 0 & 1-\lambda_m\end{matrix} \right],$
 can be used as a tool to investigate physical properties such as disorder lines\cite{MingPeschel}, Berry phase \cite{RyuHatsugai06} and  zero-energy edge states \cite{RyuHatsugai06}. While  the entanglement entropy $S_A \equiv \sum_m S_m$ where
 \be \label{Sent} S_m = -  \lambda_m \log_2{\lambda_m} - (1-\lambda_m) \log_2{(1-\lambda_m)},\ee
 contains more global information about the system, the entanglement entropy spectra offer a new method to observe the microscopic quantum phenomena. 

\begin{figure}
\center
\includegraphics[width=8.5cm]{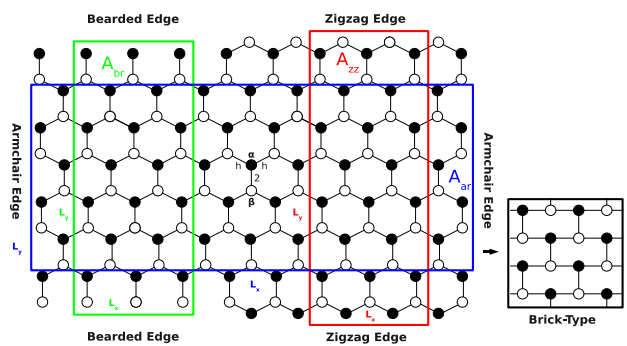}
\caption{(Color online) a honeycomb lattice with different shapes for system $A$. ($A_{br}$-green rectangle) two bearded edges, ($A_{zz}$- red rectangle) two zigzag edges,  ($A_{ar}$-blue rectangle) two armchair edges. The hopping constant along $x$-direction is $h$, along $y$-direction $2$. On the right hand side the transformation from a honeycomb lattice to a brick-type lattice is shown.      } \label{fig1}
\end{figure} 

In this Letter we explore the bipartite entanglement entropy spectra of a tight-binding honeycomb lattice for different shapes of the system $A$, as shown in Fig.~{\ref{fig1}}. We show that not only the shape of the edges but also the hopping coupling $h$ along the $x$-direction influences the Berry phase $\chi$, which value decides if the system has maximal entangled states with $\lambda_m=1/2$ or not \cite{RyuHatsugai06}. In the previous literature, only the edges with infinite size have been studied \cite{RyuHatsugai02}, while  in this Letter we analytically and numerically study the edges with finite sizes. For bearded edges, the maximal entagled states appear two-by-two only if the hopping coupling $h$ passes by critical hopping couplings $h_c$s. These results are verified numerically using the entanglement entropy spectra by diagonalizing the block Green's function matrix.

We consider a honeycomb lattice with anisotropic hopping constants as shown in Fig.~{\ref{fig1}}. The hopping coupling along the y direction is defined as $2$ of  magnitude, while the other two couplings are replaced  with a general value $h$.  Without changing the lattice topology, a honeycomb lattice can be transformed into a brick-type lattice\cite{Wakabayashi}, therefore the Hamiltonian can be rewritten as 
\be \label{Hgp}
   {\cal H} = -\sum_{x,y} \{ h c^{\dagger}_{x,y} c_{x+1,y} +\left[1+(-1)^{x+y}\right]  c^{\dagger}_{x,y} c_{x,y+1}+ h.c. \}. 
\ee
The dispersion relation for such Hamiltonian is: $\varepsilon(\vec{k}) =  2{(1+h^2 \cos^2{k_x} + 2h\cos{k_x}\cos{k_y})^{1/2}}.$ For $0<h<1$, the system is gapped, while for $h>1$ we find two Dirac Fermi points  located at $(k_x, k_y) = (\pm \cos^{-1}{(-1/h)},0)$. For $h=1$, the two Dirac zero-energy modes merge into a confluent point $(k_x,k_y) = (\pi,0)$.  The zero-energy edge states for bearded edges appear only if the bulk has two Dirac points. The reason is described as follows. 

\begin{figure}
\center
\includegraphics[width=8.5cm]{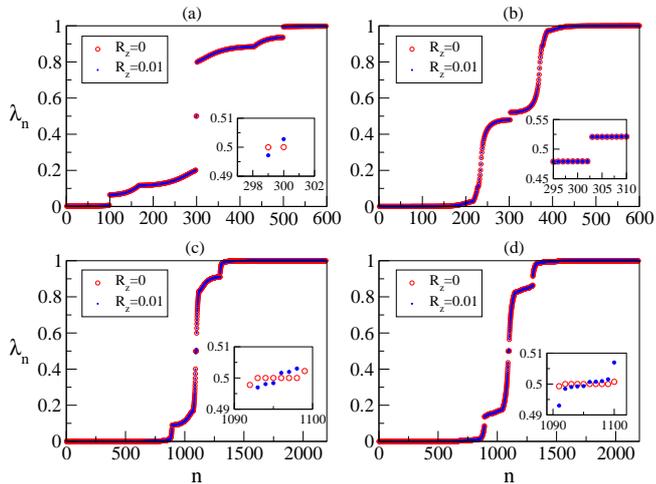}
\caption{(Color online) Entanglement entropy spectra (red circles: $R_z=0$, blue dots:$R_z=0.01$) (a) $h=2, L_x = 3$ and $L_y=200$ (two bearded edges); (b)$h=2, L_x = 201$ and  $L_y=4$ (two armchair edges); (c) $h=2, L_x=11$ and $L_y=200$ (two bearded edges); (d) $h=4, L_x=11$ and $L_y=200$ (two bearded edges). The insets show the number of maximal entangled states.} \label{fig2}
\end{figure} 

The edges on the honeycomb lattice can be produced by imposing 
different open boundaries: bearded, zigzag or armchair edges. To further see the influence of the boundaries, the sub-lattices $\gamma$ and $\delta$ can be first labeled as $\circ$ and $\bullet$ in Fig.~{\ref{fig1}}. Defining a pair of annihilation operators $\vec{c}_x^{T} = (c_{\gamma}, c_{\delta})_x$ and Fourier transforming the Hamiltonian (\ref{Hgp}), we can rewrite the Hamiltonian (\ref{Hgp}) in the momentum space as 
\be \label{Htopo}
   {\cal H} = -\sum_{\vec{k} \in BZ} \vec{c}_{\vec{k}}^{\dagger} [\vec{R}(\vec{k})\cdot \mathbf{\sigma} ]\vec{c}_{\vec{k}}, 
\ee
where $\mathbf{\sigma} = (\sigma_x,\sigma_y,\sigma_z) $ as Pauli matrices and $\vec{R}(\vec{k}) = (R_x,R_y,R_z) \in \mathbbm{R}^3$. If we consider systems without onsite potentials, one can always rotate $\vec{R(\vec{k})}$ to lie on a two-dimensional plane by applying a global $SO(3)$ unitary transformation, thus one can define  $\vec{R(\vec{k})} \equiv (R_x,R_y)$.
In this parametrization, the energy $\varepsilon(\vec{k}) = \pm {|\vec{R}|}$.  The form of  $\vec{R}(\vec{k})$ is given by the choice of the unit cell in Fourier transformation  along the  edges \cite{RyuHatsugai02}. (a) For a bearded edge along $x$-direction, $\vec{R}(\vec{k}) = (h[\cos{k_y}+\cos{(k_x-k_y)}]+2, h[\sin{k_y}-\sin{(k_x-k_y)}])$; (b) for a zigzag edge along $x$-direction,  $\vec{R}(\vec{k}) = (2\cos{(k_x-k_y)} + h[1+\cos{k_x}], h\sin{k_x}-2\sin{(k_x-k_y)})$; (c) for a armchair edge along $y$ direction,  $\vec{R}(\vec{k}) = (h[\cos{k_x}+\cos{(k_x+k_y)}]+2, h[\sin{k_x}+\sin{(k_x+k_y)}])$. The choice of the unit cell in Fourier transformation is not unique, however, different choices lead to  the same topology for the edge states \cite{Footnote}. By fixing  a wave factor parallel to  the edges and investigating  the loop of $\vec{R}$ as a parameter of the perpendicular wave vector changing from $-\pi$ to $\pi$, the topology of the system can be obtained. In case that the loop $\ell$ of $\vec{R}$ contains the origin $\cal{O}$ in the $\vec{R}$ space, the Berry phase (or Zak's phase) $\chi$, defined as a line integral of the curvature of the filled band, is $\pi$. In this case, due to the fact that one can  continuously deform $\ell$ into a unit circle without crossing the origin,  the topological argument ensures us that the original Hamiltonian corresponding to $\ell$ contains at least one zero-energy edge state \cite{RyuHatsugai02}.  Therefore for (a) bearded edges along $x$-direction, where $k_x$ is fixed and $k_y$ serves as parameter of the loops, there exists a zero energy state only if $-2\cos^{-1}{(1/h)} \leq k_x \leq2\cos^{-1}{(1/h)} $. Similar situation happens  for (b) zig-zag edges. The zero-energy edges states appear under the condition that $-\pi \leq k_x \leq -\cos^{-1}{(2/h^2-1)}$ or $\cos^{-1}{(2/h^2-1)} \leq k_x \leq \pi$. On the other hand, for (c) armchair edges, no zero-energy states exist due to the fact that no loops will encircle $\cal{O}$ with fixed  $k_y$ by changing $k_x$. Therefore for the zigzag and bearded edges the hoping coupling $h$ changes the range of $k_x$ where  the Berry phase $\chi = \pi$.

The consideration for entanglement entropy is somehow different: we no longer cut the system with an open boundary, but partition that into two parts: system $A$ and environment $B$. By using von Neumann entropy $S_A$, one can figure out how the system $A$ entangles with the environment $B$.   However, there exists a one-to-one correspondence between the zero-energy state for the Hamiltonian with edges and the maximal entangled state for the bipartite system. The reason is as follows. Consider the whole bipartite $AB$ consisting of  $N$ sites  (or modes), with $n$ sites of system $A$ and notice that  in our calculation only thermodynamic limit ($N \rightarrow \infty$) will be taken. The reduced density matrix $\rho_A = \tr_B \rho_0 $ where $\rho_0\equiv|\Psi_{AB}\rangle\langle\Psi_{AB}| $  can be obtained by determining the matrix elements of the full density matrix with respect to coherent states and integrating out the variables of the environment $B$ \cite{MingPeschel} and can be related to the eigenvalues $\lambda_m$ of the block Green's function matrix\cite{HenleyPeschelMing} $G_{\gamma\delta}(\vec{r}_i-\vec{r}_j) = \tr \rho c_{(\vec{r}_i,\gamma)} c_{(\vec{r}_j,\delta)}^{\dagger}$, where $(\vec{r_i},\gamma)$ and $(\vec{r}_j,\delta)$ belong to system $A$. The entanglement entropy therefore takes the form as Eq. (\ref{Sent}). In this case, the Green's function matrix can be Fourier transformed as $G_{\gamma\delta}(\vec{r}_i-\vec{r}_j) = N^{-1} \sum_{\vec{k} \in BZ} e^{-i \vec{k}\cdot(\vec{\vec{r}_i-\vec{r}_j})}G_{\gamma\delta}(\vec{k})$, where $ G_{\gamma\delta}(\vec{k})$ is calculated as  
\be \label{GTopo}
    G_{\gamma\delta}(\vec{k}) = \frac{1}{2}  \left[ 1 - \frac{\vec{R} \cdot \vec{\sigma}}{R} \right]_{\gamma\delta}.
\ee  
We regard  $G_{\gamma\delta}(\vec{r}_i-\vec{r}_j)$ as an effective Hamiltonian for obtaining the entanglement entropy. In the case of taking the whole infinite plane with periodic boundary, i.e. no sites in the environment $B$, the $G$ Hamiltonian has the same set of eigenfunctions as the original Hamiltonian, however, the eigenvalues are either $1$ or $0$, which means $S=0$, the system is not entangled at all.  On the other hand, while $B$ is not empty, the nontrivial boundary states appear, which eigenvalues are not $1$ or $0$. This boundary states, which eigenvalues $0<\lambda_m<1$, cause  the nonzero entanglement entropy, therefore we can say that those  are the most important states for calculating the entanglement.  

Comparing the Hamiltonian (\ref{Htopo}) with $G_{\gamma\delta}(\vec{k})$ (\ref{GTopo}), they almost take the same form except a constant and a positive normalization factor $R$, therefore we can conclude that they should share the same topology, that is, if $\vec{R}$ encloses origin in the parameter space which causes  Berry phase equal to $\pi$, the zero-energy state appears for the Hamiltonian (\ref{Htopo}), while  for the block Green's function matrix we obtain one special  eigenvalue $\lambda_m = 1/2$ \cite{RyuHatsugai06}. They also share the same eigenfunctions with different eigenvalues. We call the state with $\lambda_m =1/2$ a maximal entangled state due to the fact that the resulting $S_m = 1$. According to the discussion above, only the reduced density matrix of  the system $A$ with bearded and zigzag edges has maximal entangled eigenstates, while for the armchair edges there exist no maximal entangled states. 

\begin{figure}
\center
\includegraphics[width=8cm]{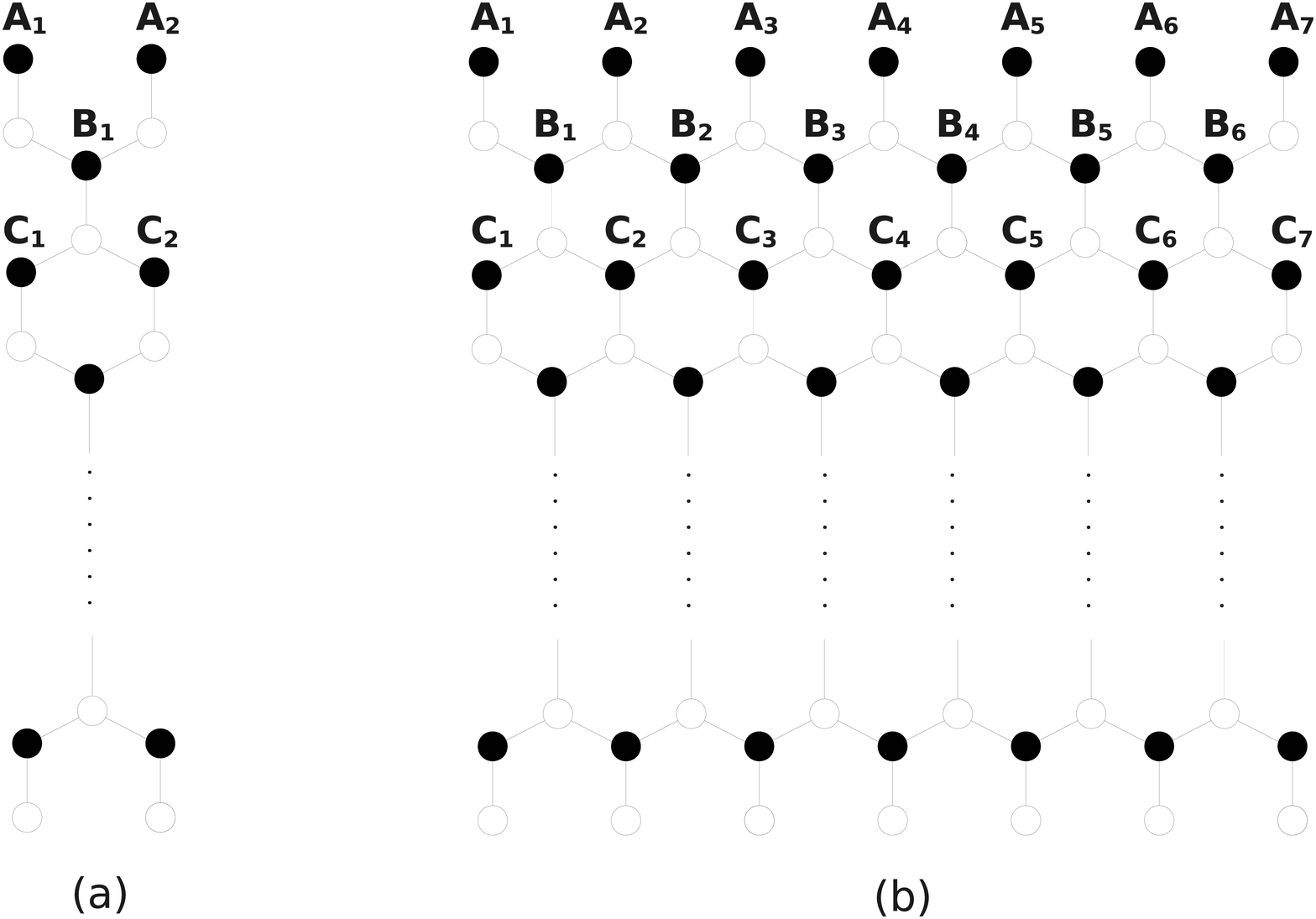}
\caption{Position labeling for wavefunctions of the honeycomb lattice. (a)$L_x=3$ (b) $L_x=11$.} \label{fig3}
\end{figure}

In order to investigate this effect we numerically diagonalize the block Green's function matrix  $G_{\gamma\delta}(\vec{r}_i-\vec{r}_j)$ for the system $A$ with finite size, as shown in Fig.~{\ref{fig1}}, which $L_x$ and $L_y$ are defined as follows. Transforming the honeycomb lattice into the brick-type lattice and projecting the system sites into $x$ and $y$ direction, $L_x(L_y)$ is defined as the total number of the projected sites along the $x(y)$ direction. 
 The real noninteracting edge state only appears in the thermodynamic limit, i.e. $L_x \gg L_y$ or vice versa. For example the partition of system $A_{br}$ and $A_{zz}$in Fig.~{\ref{fig1}} have either two bearded $(A_{br})$ or zigzag edges $(A_{zz})$. On the other hand there exist two armchair edges for the system $A_{ar}$. Here we only consider the edges far apart from each other so that the two edges do not interact with each other (independent edges). Therefore we expect that for the partitions $A_{br}$ or $A_{zz}$ there exist maximal entangled states $(\lambda_m=1/2)$, while for $A_{ar}$ the maximal entangled states do not appear. (Due to the fact that bearded and zigzag edges have the same properties, we will only calculate the entanglement spectra for the bearded edges in the following.)    
The upper panels of Fig.~\ref{fig2} shows the entanglement spectra $\lambda_m$ of a system size: (a) $L_x=3, L_y=200$ ($A_{br}$-type) and (b) $L_x=201, L_y =4$ ($A_{ar}$-type)  for a graphene $(h=2)$. For the system with two independent bearded edges, we obtain two maximal entangled states with $\lambda_m=1/2$ (see the inset), while there are no maximal entangles states for the system $L_x \gg L_y$ (on the right hand side of Fig.~\ref{fig1}) due to the fact that $A_{ar}$-type system has two independent armchair edges. Fig.~{\ref{fig2}} (c) shows the spectra for $L_x=11, L_y=200$ and $h=2$. From the inset we observe $6$ eigenvalues close to $1/2$. In Fig.~{\ref{fig2}}(d) the system size is the same as (c) but the hopping coupling $h=4$. We obtain $10$ maximal entangled states (Due to finite-size effect, two eigenvalues of the states are not precise $1/2$, however we still count them as maximal entangled states). The number of maximal entangled states not only depend on the shape of the system but also on the hopping coupling $h$. We will discuss that as follows.   

For general $h$, the wave function of the maximal entangled state is the same as the zero-energy mode. The zero-energy wavefunction is obtained by the tight-binding Schr\"odinger equation (SE)
$\label{SE} -\sum_j t_j \psi_j = E \psi_i =0,$ 
where $j$ are neighbor sites of $i$, $t_j=h$ for the vertical hopping couplings and $t_j=2$ for the longitudinal hopping constants in the brick-type representation. 
Fig.~{\ref{fig3}} (a) shows the structure of the system with $L_y \gg L_x = 3$.  To obey SE, a solution with all values of the wavefunction for $\circ$ sites in Fig.~{\ref{fig3}} equal to zero is expected. For each partition  symmetric and antisymmetric wavefunctions with respect to the middle line of the system in the  $x$ direction can be separated to two different groups. For symmetric solutions, we set that $[\psi_{A_1}, \psi_{A_2}] = [\phi,\phi]$. According to SE, $\psi_{B_1} = -(2/h) \psi$ and  $[\psi_{C_1}, \psi_{C_2}] = [(2/h^2)\psi,(2/h^2) \phi]$. Therefore there only exist maximal entangled states (zero-energy states) if $2/h^2 <1$, otherwise the wavefunction will explode with the increasing $L_y$. We obtain a critical $h_c = \sqrt{2}$:  the maximal entangled states appear only if $h \leq h_c$.  On the other hand the antisymmetric solutions $[\psi_{A_1}, \psi_{A_2}] = [\phi,-\phi]$ fail for $L_x=3$ due to the fact that the value on site $B_1$ directly sets  $\phi=0$. 

Another example we study here is $L_x=11$, as shown in Fig.~{\ref{fig3}} (b). For the symmetric wavefunctions, we can first set the values $[\psi_{A_i}] =  [1, \alpha_1, \alpha_2,\alpha_2,\alpha_1,1] \phi$ and $[\psi_{B_i}] = [\beta_1,\beta_2,\beta_3,\beta_2,\beta_1] \phi$. Due to the fact that the structure of the layer $C_i$ is the same as $A_i$, the values of the wavefunction can be set as $[\psi_{C_i}] = x [\psi_{A_i}] $. To obtain the critical couplings $h_c$s we have to know all these values $\alpha_i, \beta_i$ and $x$ by using SE in different $\circ$ sites. The SE for  the site between $A_1$ and $B_1$ gives $\beta_1 = -2/h$. The site among $A_2, B_1$ and $B_2$ leads to the equation $\beta_2 = 2(1-\alpha_1)/h$. In the same manner  we obtain $\beta_3 = 2(\alpha_2-\alpha_1-1)/h$. Using these relations and the SE for the $\circ$ site between $B_2$ and $C_2$, the first value for $x$ is found: $x=4/[(1+\alpha_1)h^2]$. From this relation the critical coupling is obtained by setting $x=1$, i.e.
 $\label{hc}h_c = 2/\sqrt{1+\alpha_1}.$
 $h_c$ is therefore obtained if $\alpha_1$ is known. On the other hand, different relations for $x$ can be obtained by subsequently calculating the SE for the $\circ$ sites among $B_i$  and $C_i$: $x=4(\alpha_1-1)/[(\alpha_1+\alpha_2)h^2]$ and $x=2(1+\alpha_2-\alpha_1)/(\alpha_2 h^2)$. These three $x$ should be identical, we first obtain  $\alpha_2 = \alpha_1^2-\alpha_1-1$ and finally the polynomials for $\alpha_1$ is found : $\alpha_1^3 - 3\alpha_1^2 +2 =0$. This indicates that $\alpha_1=1$ or $\alpha_1 = 1\pm \sqrt{3}$. For the symmetric eigenfunctions we obtain three critical couplings $h_c = \sqrt{2},  2\sqrt{2-\sqrt{3}}$ and $2\sqrt{2+\sqrt{3}} $ corresponding to different $\alpha_1$.

\begin{figure}
\vspace{1cm}
\center
\includegraphics[width=8.5cm]{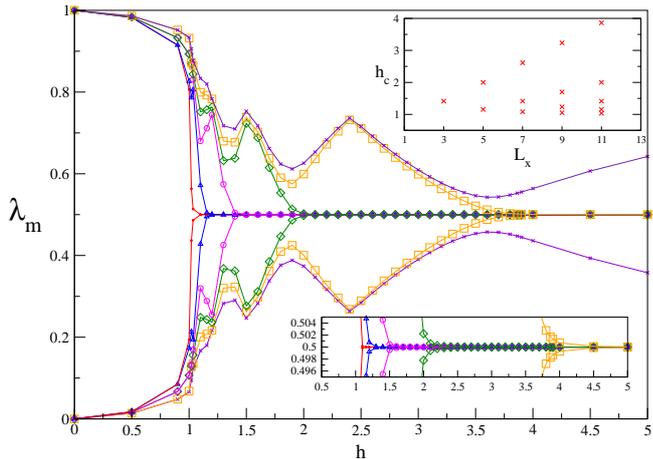}
\caption{(Color online) (The main figure) Numerical result of the nearest eigenvalues close to $1/2$ with varying $h$ for $L_x=11$ and $L_y=200$. (Lower inset) Zooming for $\lambda_m$ near $1/2$. There exist five critical couplings $h_c$. (Upper inset) Analytical result for $h_c$ with different sizes: $L_x=3,5,7,9,11$.} \label{fig4}
\end{figure} 

 The antisymmetric eigenfunctions  can be also found in a similar way. We can first assume that $[\psi_{A_i}] =   [1, \alpha_1, \alpha_2,-\alpha_2,-\alpha_1,-1] \phi$, $[\psi_{B_i}] = [\beta_1,\beta_2,0,-\beta_2,-\beta_1] \phi$ and $[\psi_{C_i}] = x [\psi_{A_i}]$. From the layer between $A_i$ and $B_i$ one obtains $\beta_1= -2/h$ and $\beta_2=2(1-\alpha_2)/h$. The SE  for the site among $A_3$, $B_2$ and $B_3$ gives the relation as $\alpha_2 = (\alpha_1-1)$. Two relations for $x$ can be obtained by solving the SE for the sites between $B_i$ and $C_i$: $x= 4/[(1+\alpha_1)h^2]$ and $x=4(\alpha_1-1)/[(\alpha_1+\alpha_2)h^2]$. Using these relations we finally obtain $\alpha_1 =0,2$ and $h_c = 2\sqrt{3}/3$ and $2$. Fig.~{\ref{fig4}} shows the some eigenvalues $\lambda_m$ nearest $0.5$ of the system $L_x=11$ and $L_y=200$ with bearded edges along $x$-direction as a function of $h$ by diagonalizing the block Green's function matrix. The lower inset shows that the system has $5$ different $h_c$s and they coincide with  the analytical values we discussed above.

The upper inset in Fig.~{\ref{fig4}} shows the analytical values of $h_c$s with varied $L_x$.   Firstly, for a fix $h$, the eigenvalues increase as a function of $L_x$. In the case of $L_x \rightarrow \infty$, those eigenvalues  form a continuum, i.e. form a flat band of zero surface states, as shown in \cite{Wakabayashi} for   graphenes ($h=2$). The number of eigenvalues can be approximated as $N \approx 2/\pi L_x \cos^{-1}{(1/h)}$ for large $L_x$. This accounts for the increasing number of the eigenvalues as $h$  increases. 

Secondly we found that there are some repeated $h_c$s  with growing $L_x$. For example: for $L_x=3, 7, 11$, the system has the same $h_c = \sqrt{2}$. The reason is that one can build either a symmetric or antisymmetric wavefunction for $L_x=7$ and $11$ using a fundamental block $L_x=3$ similar to the toy Lego : a positive and a negative block of $L_x=3$ can build a system with $L_x=7$; a sandwich-like system with two $L_x=3$ positive blocks on both sides and a negative block in the middle  has exact the size $L_x=11$. Similarly, for $L_x=5$ and $L_x=11$ there exists the same $h_c=2$. It is clear that a positive and a negative block with $L_x=5$ can build a system with $L_x=11$. 

An interesting question arises: if considering a system with onsite potentials, i.e. if putting  $\vec{R} = (R_x, R_y, R_z), R_z= \Delta$, do the edges states still robustly exist? By adding the onsite potentials, the chiral symmetry is destroyed, there exist no maximal entangled  states. The blue lines and dots of the inset  in Fig.\ref{fig2} shows the small deviation of $\lambda=1/2$ by putting $\Delta=0.01$. However, due to the fact that the deviation is small, we can conclude that the states with eigenvalues closed to $1/2$ are still edge states. That means, the edge states robustly exist even with broken chiral symmetry.

In the summary, we have  used the entanglement spectra to study the edge states of a honeycomb lattice. The dependence between the Berry phase and hopping couplings has been found. The one-to-one correspondence between the maximal entangled state for a bipartite system and the zero energy state has been proved numerically by diagonalizing the block Green's function matrix. We found there exist critical coupling couplings $h_c$s for the system with finite edges: the maximal entangled states pair-by-pair  increases every time when the $h$ jumps over $h_c$s.

\bigskip

%%%%%%%%%%%%%%%%%%%%%%%%%%%%%%%%%%%%%%%%%%%%%%%%%%%%
%\newpage


\begin{thebibliography}{30}
\bibitem{Footnote} We choose non-orthogonal $k_x$ and  $k_y$ according to different choice of the unit cell in the Fourier transformation. In order to obtain the same dispersion relation for the Hamiltonian (\ref{Hgp}), a linear transformation is needed. For example for bearded edges: $k_x \rightarrow 2k_x$ and $k_y \rightarrow k_y + k_x$.
\end{thebibliography}
\end{document}